\documentclass[aps,twocolumn,PRL,reprint,superscriptaddress]{revtex4}
\usepackage{amsmath}
\usepackage{times}
\usepackage{graphicx}
\usepackage{bm}
\usepackage{amssymb}
\usepackage{hyperref}
\newcommand{\bi}[1]{{\boldsymbol{#1}}}
\begin{document}
\title{Topological semimetals with double-helix nodal link}
\author{Wei Chen}
\affiliation{College of Science, Nanjing University of Aeronautics and Astronautics, Nanjing 210016, China}
\author{Hai-Zhou Lu}
\affiliation{Institute for Quantum Science and Engineering and Department of Physics, South University of Science and Technology of China, Shenzhen 518055, China}
\author{Jing-Min Hou}
\email{Corresponding author: jmhou@seu.edu.cn}
\affiliation{School of Physics, Southeast University, Nanjing 211189, China}
\begin{abstract}
Topological nodal line semimetals are characterized by the crossing of the conduction and valence bands along one or more closed loops in the Brillouin zone. Usually, these loops are either isolated or touch each other at some highly symmetric points. Here, we introduce a new kind of nodal line semimetal, that contains a pair of linked nodal loops. A concrete two-band model was constructed, which supports a pair of nodal lines with a double-helix structure, which can be further twisted into a Hopf link because of the periodicity of the Brillouin zone. The nodal lines are stabilized by the combined spatial inversion $\mathcal{P}$ and time reversal $\mathcal{T}$ symmetry; the individual $\mathcal{P}$ and $\mathcal{T}$ symmetries must be broken. The band exhibits nontrivial topology that each nodal loop carries a $\pi$ Berry flux. Surface flat bands emerge at the open boundary and are exactly encircled by the projection of the nodal lines on the surface Brillouin zone. The experimental implementation of our model using cold atoms in optical lattices is discussed.
\end{abstract}

\maketitle

\emph{Introduction.}---The recent discovery of topological insulators and superconductors has greatly deepened our understanding of the quantum phases of matter \cite{Hasan,Qi}. For a gapped phase, the band topology can be well-described using topological invariants in terms of symmetries \cite{Qi2,Schnyder,Kitaev,Chiu,Slager,Kruthoff}. As the conduction and valence bands cross each other in the Brillouin zone, the system enters a semimetal phase. The topology of the gapless phase brings totally different stories, which gives rise to the concept of topological semimetals \cite{Weng}. In three dimensions, the band crossing that carries nontrivial topology may occur at discrete points or along closed loops. The former case corresponds to Weyl/Dirac semimetals \cite{Murakami,Wan,Wang1,Wang2}, whereas the latter case corresponds to topological nodal line semimetals (TNLSMs) \cite{Burkov}. Weyl and Dirac semimetals \cite{Weng2,Huang,Lv,Xu2,Lv2,XHuang,Lv3,Liu,Xu,Xiong,Liu2,Borisenko,Yi} have both been experimentally confirmed, which has increased research interest in topological semimetals. Now the latest member of topological semimetals, TNLSM, is waiting for experimental verifications \cite{Fang}. A variety  of candidates have been proposed \cite{Volovik,Weng3,Chen,Zeng,Kim,Yu,Fang2,Yamakage,Xie,Chan,Bian,Zhao}, and recent experiments have shown promising results based on angle-resolved photoemission \cite{Bian} and quantum oscillations \cite{Hu,JPLiu} measurements. Nodal lines have been shown to play a key role in topological field theories in the Brillouin zone \cite{Lian}.

\begin{figure}
\centering
\includegraphics[width=0.45\textwidth]{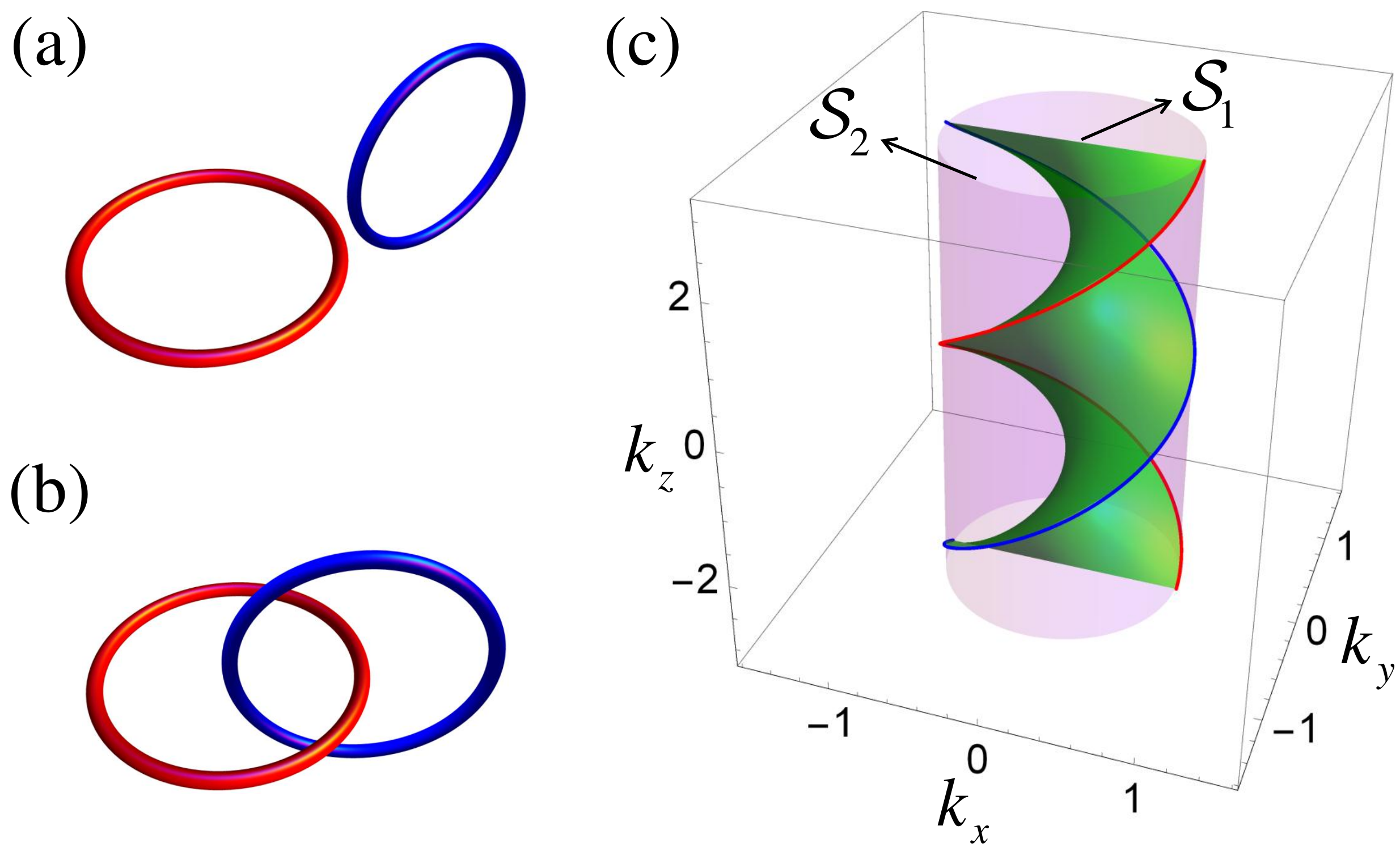}
\caption{(Color online). Topological configurations of two closed nodal loops: (a) isolated, or (b) forming a Hopf link. (c) Nodal lines with a double-helix structure, formed by intersecting lines of two surfaces $\mathcal{S}_1$ and $\mathcal{S}_2$.
 } \label{fig1}
\end{figure}

In addition, the topological classification of TNLSMs remains incomplete \cite{Chiu,Fang}. Unlike the topology of Weyl and Dirac semimetals, which can be well-described by a single topological invariant under proper symmetry protections, the topology of TNLSM is more subtle \cite{Fang}. From a simple geometrical perspective, there are two configurations of zero dimensional nodal points, in which they are either coincident or not. In contrast, there are a variety of possible configurations for one dimensional nodal loops. They can (i) be isolated (Fig. \ref{fig1}(a)), (ii) touch at certain points, or (iii) be linked with each other (Fig. \ref{fig1}(b)). This intrinsic difference may considerably enrich the scenarios of TNLSMs. A typical example of the first case is the system described by the minimal model $h(\bm{k})=\sin k_z\sigma_x+[M-4B(\sin^2\frac{k_x}{2}+\sin^2\frac{k_y}{2}+\sin^2\frac{k_z}{2})]\sigma_z$, where two isolated nodal loops lie in two planes $k_z=0,\pi$. The second case has also been reported; for instance, gyroscope-shaped nodal lines \cite{Kim,Yu} and the recently predicted non-symmorphic nodal chain metals \cite{Chain}.

In this Rapid Communication, we determined the existence of a third kind of TNLSM, in which nodal loops are linked. In contrast to existing TNLSMs, a pair of nodal lines pass through each other and form a double-helix structure, as shown in Fig. \ref{fig1}(c). Because of the periodicity of the Brillouin zone, this double-helix is topologically equivalent to a Hopf link, carrying a nonzero linking number. Moreover, the band also possesses nontrivial topology, in which each nodal line carries a $\pi$ Berry flux, which results in novel surface states.

\emph{Two-band model with double-helix nodal link.}---We constructed a tight binding model based on a cubic lattice. This model can be described in momentum space as follows:

\begin{equation}\label{H}
\begin{split}
\mathcal{H}(\bm{k})&=d_x(\bm{k})\sigma_x-d_z(\bm{k})\sigma_z,\\
d_x(\bm{k})&=\sin k_y\cos k_z-\sin k_x \sin k_z,\\
d_z(\bm{k})&=2\cos k_x+2\cos k_y+\chi,
\end{split}
\end{equation}
where the Hamiltonian is chosen to be dimensionless for simplicity, the lattice constant is set to unity, $\sigma_{x,y,z}$ are the Pauli matrices for pseudospin (such as the orbital degree of freedom), and $\chi$ is a tunable parameter. Our model can also be formulated with the Hopf map method by designing a specific mapping form \cite{Moore,Deng1,Deng2,Kennedy,CXu} (see Supplemental Material \cite{SM}).

Diagonalizing the Hamiltonian produces the eigenstates $|u_\pm(\bm{k})\rangle$ of the Hamiltonian (\ref{H}) of opposite energies $E_\pm(\bm{k})=\pm\sqrt{d_x^2+d_z^2}$. Degeneracy of the bands occurs when $d_x(\bm{k})=d_z(\bm{k})=0$, which defines the nodal lines in the Brillouin zone. It can be interpreted as the intersecting lines of two surfaces $\mathcal{S}_1:d_x(\bm{k})=0$ and $\mathcal{S}_2:d_z(\bm{k})=0$. Here, the nodal lines form a novel double-helix structure. This becomes explicit at the limit of $k_x\ll1, k_y\ll1$, in which the parametric equation of $\mathcal{S}_1$ reduces to $k_y/k_x=\tan k_z$, which describes a helicoid, while that of $\mathcal{S}_2$ becomes $k_x^2+k_y^2=4+\chi\ll1$, corresponding to a cylinder. The intersecting lines of the two surfaces pass through each other to form a double-helix, as shown in Fig. \ref{fig1}(c). Moreover, because of the periodicity of the Brillouin zone, the cylinder $\mathcal{S}_2$ folds into a torus. Correspondingly, the double-helix structure folds into a Hopf link (Fig. \ref{fig1}(b)). Such a ``double-helix nodal link'' (DHNL) possesses a nonzero linking number, so the nodal loops cannot shrink to a point without crossing each other.

\emph{Symmetry analysis}.---In general, the nodal lines in TNLSMs are stabilized by extra symmetries imposed on the Hamiltonian \cite{Burkov}. This situation should also be true for our model. For a system without spin-orbit coupling, the time reversal operator $\mathcal{T}$ acts on the Hamiltonian through $\mathcal{T}\mathcal{H}(\bm{k})\mathcal{T}^{-1}=\mathcal{H}^*(-\bm{k})$, and the spatial inversion operator $\mathcal{P}$ reverses the momentum as $\mathcal{P}\mathcal{H}(\bm{k})\mathcal{P}^{-1}=\mathcal{H}(-\bm{k})$. The semimetal phase with DHNL breaks both $\mathcal{T}$ and $\mathcal{P}$ symmetries, as follows:
\begin{equation}
[\mathcal{H},\mathcal{T}]\neq 0, [\mathcal{H},\mathcal{P}]\neq 0.
\end{equation}
Moreover, it does not have reflection symmetry $\mathcal{R}$ about any plane. Thus, our scheme differs from those of existing TNLSMs, in which at least one of these three symmetries exists \cite{Weng3,Chen,Zeng,Kim,Yu,Fang2,Yamakage,Xie,Chan,Bian,Zhao}. However, without individual $\mathcal{P, T, R}$ symmetries, the system retains combined $\mathcal{PT}$ symmetry \cite{Zhao2,Zhang} as
\begin{equation}\label{pt}
[\mathcal{H}(\bm{k}),\mathcal{PT}]=\mathcal{H}(\bm{k}),
\end{equation}
which reflects the reality of the Hamiltonian. In addition, the Hamiltonian (\ref{H}) respects chiral symmetry, which can be described by the anticommutation relation as
\begin{equation}\label{chiral}
\{\mathcal{H}(\bm{k}),\Gamma\}=0,
\end{equation}
where the chiral operator $\Gamma=i\sigma_y$ corresponds to twofold spin rotation. The $\Gamma$ symmetry guarantees that the eigenstates $|u_\pm(\bm{k})\rangle$ with opposite energies always exist in pairs. Both symmetries forbid the $d_y(\bm{k})\sigma_y$ term to enter $\mathcal{H}(\bm{k})$, which is essential for the stability of the DHNL. The difference between these two symmetries is that the $\Gamma$ symmetry simultaneously excludes the energy term $d_0(\bm{k})\sigma_0$ (where $\sigma_0$ is the unit matrix), whereas the $\mathcal{PT}$ symmetry does not. As a result, the chiral symmetry not only stabilizes the DHNL, it also restricts its energy to zero.

\begin{figure*}
\centering
\includegraphics[width=0.7\textwidth]{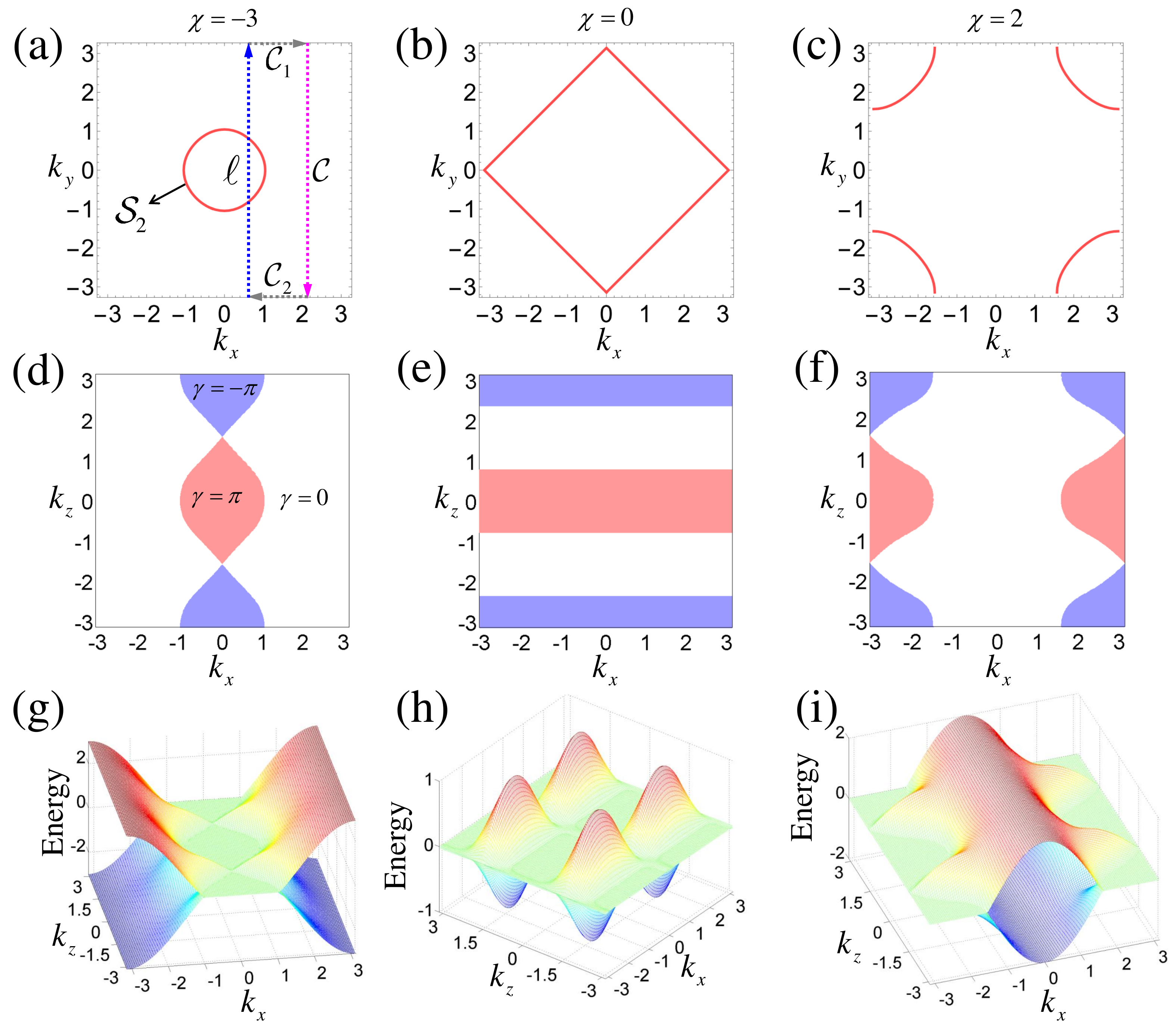}
\caption{(Color online). (a)-(c) Projection of the surface $\mathcal{S}_2$ with different values of $\chi$ (circle for $\chi<0$, square for $\chi=0$, and circle centered around $k_x=k_y=\pi$ for $\chi>0$) in the $k_x-k_y$ plane. The integral paths $\ell\rightarrow\mathcal{C}_1\rightarrow\mathcal{C}\rightarrow\mathcal{C}_2$ are indicated by the arrowed lines. (d)-(f) Berry phase distribution in the projected surface Brillouin zone as a function of $k_x$ and $k_z$, with values of $\chi$ corresponding to those in (a)-(c), respectively. (g)-(i) Surface states at open boundaries in the $y$ direction.} \label{fig2}
\end{figure*}

\emph{Band topology and surface states}.---In addition to its novel geometric configuration, the DHNL also exhibits nontrivial topology. Each of the two nodal lines of the DHNL carries a $\pi$ Berry flux in the Brillouin zone. Therefore, if one were to travel along any closed path encircling one of the nodal lines, the accumulated geometric phase would be equal to $\pi$. To demonstrate this, we first prove the stability condition for the DHNL, that is, the Berry curvature $\bm{\Omega}(\bm{k})$ generically vanishes for nondegenerate points \cite{Burkov}. Away from the DHNL, the Berry curvature for the valence band can be defined as
\begin{equation}
\Omega_\mu^-(\bm{k})=i\epsilon_{\mu\nu\lambda}\langle\partial_\nu u_-(\bm{k})|\partial_\lambda u_-(\bm{k})\rangle,
\end{equation}
where $|u_-(\bm{k})\rangle$ is the periodic part of the Bloch wave function, $\partial_\nu$ denotes $\partial/\partial k_\nu$, and $\epsilon_{\mu\nu\lambda}$ is the Levi-Civit\`{a} antisymmetric tensor. Due to the chiral symmetry (\ref{chiral}), if $|u_-(\bm{k})\rangle$ is the wave function for the valence band, $|u_+(\bm{k})\rangle=\sigma_y|u_-(\bm{k})\rangle$ must be the eigenstate of the conduction band. A direct calculation shows that the Berry curvature of the valence band is equal to that of the conduction band for a given $\bm{k}$, $\bm{\Omega}^-(\bm{k})=\bm{\Omega}^+(\bm{k})$. By using the local conservation law for the Berry curvature \cite{Xiao} $\bm{\Omega}^-(\bm{k})+\bm{\Omega}^+(\bm{k})=0$, we obtain
\begin{equation}
\bm{\Omega}^-(\bm{k})=\bm{\Omega}^+(\bm{k})=0.
\end{equation}
This indicates that if a nonzero distribution of the Berry curvature exists in the Brillouin zone, it must be strictly localized along the DHNL. Then we may choose an arbitrary integral loop enclosing a single nodal line to calculate the Berry phase.

Without loss of generality, we can choose the integral loop of $\ell\rightarrow\mathcal{C}_1\rightarrow\mathcal{C}\rightarrow\mathcal{C}_2$, as indicated by the arrowed lines in Fig. \ref{fig2}(a). This loop is composed of two line segments parallel to the $y$ axis ($\ell$ and $\mathcal{C}$) and two segments parallel to the $x$ axis ($\mathcal{C}_1$ and $\mathcal{C}_2$). The two $x$-axis paths $\mathcal{C}_1$ and $\mathcal{C}_2$ are equivalent, but oriented in opposite directions; therefore, their contributions cancel out \cite{Abanin}. Such a scheme benefits subsequent discussion on the surface states. We can regard $k_x$ and $k_z$ as parameters and derive the Berry phase of the effective one dimensional model along the $y$ direction. The wave function of the valence band is $|u_-(\bm{k})\rangle=\Big(\text{sgn}(d_x)\sqrt{\frac{1-\cos\theta}{2}},-\sqrt{\frac{1+\cos\theta}{2}}\Big)^T$, where $\cos\theta=-d_z/\sqrt{d_x^2+d_z^2}$. The Berry phase can be obtained as \cite{Shen}
\begin{eqnarray}
\gamma&=&i\int_{-\pi}^\pi dk_y\langle u_-(\bm{k})|\partial_{k_y}|u_-(\bm{k})\rangle\nonumber\\
&=&\frac{i}{2}\int_{-\pi}^\pi dk_y\big[\partial_{k_y}\ln \text{sgn}(d_x)\big](1-\cos\theta),\label{B1}
\end{eqnarray}
in which the integral path $\ell$ is parallel to the $y$ axis, as shown in Fig. \ref{fig2}(a). A nonzero contribution to the integral may come from the $k_y$-dependent sign change of $d_x(\bm{k})$, so the integral can be written in the neighborhood of two real roots $k_y=k_1,k_2$ of the equation $\sin k_y=\sin k_x\tan k_z$. In this case, we have $\cos\theta|_{k_y=k_{1,2}}=-\text{sgn}(d_z)$, and Eq. (\ref{B1}) reduces to
\begin{eqnarray}
\gamma=\frac{i}{2}\sum_{j=1,2}\int_{k_j-\delta}^{k_j+\delta}dk_y\big[\partial_{k_y}\ln \text{sgn}(d_x)\big]\nonumber\\
\times\big[1+\text{sgn}(2\cos k_x+2\cos k_j+\chi)\big].\label{B2}
\end{eqnarray}
For given values of $k_x$ and $k_z$, the real roots $k_{1,2}$ correspond to the crossing points of $\ell$ and $\mathcal{S}_1$. Once $\ell$ passes through $\mathcal{S}_1$, it contributes a $\pi$ phase to the integral. Because $k_{1,2}$ must either exist as a pair or not at all, $\ell$ may pass through $\mathcal{S}_1$ either twice or never, and thus, the Berry phase is zero if there are no additional constraints. The second factor in the integral incorporates an additional constraint, $2\cos k_x+2\cos k_j+\chi>0$, which ensures that the integral is inside $\mathcal{S}_2$. Under this restriction, the path $\ell$ may cross $\mathcal{S}_1$ only once, resulting in a nontrivial Berry phase.

From this geometric viewpoint, it can be seen that the configuration of $\mathcal{S}_1$ and $\mathcal{S}_2$ determines the Berry phase distribution in the $k_x-k_z$ plane. When projected to the surface Brillouin zone, the DHNL defines the boundaries between the topologically trivial ($\gamma=0$) and nontrivial ($\gamma=\pm\pi$) regions. This can be verified by the numerical results of Eq. (\ref{B2}), as shown in Fig. \ref{fig2}(a)-(f). For $\chi<0$, $\mathcal{S}_2$ is a closed cylinder with a central axis located at $k_x=k_y=0$ (Fig. \ref{fig2}(a)). The area with nontrivial topology is encircled by the projected DHNL on the surface Brillouin zone (Fig. \ref{fig2}(d)). Because of its double-helix structure, the neighboring nontrivial regions have opposite Berry phases. As $\chi$ increases, the area surrounded by $\mathcal{S}_2$ expands, as does the topologically nontrivial area. At the critical point, $\chi=0$, the surface $\mathcal{S}_2$ meets the Brillouin zone boundary (Fig. \ref{fig2}(b)). At this point, the topologically nontrivial area reaches its maximum, which occupies half of the Brillouin zone (Fig. \ref{fig2}(e)). If $\chi$ increases further, $\mathcal{S}_2$ is opened, as shown in Fig. \ref{fig2}(c). In this case, the area defined by $2\cos k_x+2\cos k_j+\chi>0$ is surrounded by both $\mathcal{S}_2$ and part of the Brillouin zone boundaries. For $|k_x|<\cos^{-1}(1-\chi/2)$, there is no constraint on the interval of the integration in Eq. (\ref{B2}). Therefore, the Berry phase equals zero in this region, as shown in Fig. \ref{fig2}(f). This is equivalent to a cylinder centered at $k_x=k_y=\pi$, at which the nodal lines are confined. Considering that the topologically trivial and nontrivial regions are separated by the projection of the DHNL, we may choose the integral paths $\ell$ and $\mathcal{C}$, located on opposite sides of the DHNL, such that the integral exactly equals the $\pi$ Berry flux of each nodal line. This demonstrates the nontrivial band topology of the semimetal phase.

The nontrivial topology of the DHNL suggests the existence of surface flat bands in the projected surface Brillouin zone \cite{Burkov}. By performing a Fourier transformation on the Bloch Hamiltonian (\ref{H}) in the $y$ direction, the energy bands with open boundaries in the $y$ direction can be calculated. The two bands closest to zero energy are shown in Fig. \ref{fig2}(g-i) for different values of $\chi$. The flat bands coincide with the regions where $\gamma=\pm\pi$, suggesting that these zero modes are topologically protected. We also calculated the Berry phase integral along the $x$ direction, and its distribution was found to match the surface flat bands \cite{SM}. Because this system does not have $C_4$ rotational symmetry, the Berry phase distribution calculated along the $x$ and $y$ directions are different. Moreover, by introducing an additional term into $d_x$ in Eq. (\ref{H}), the DHNL can be unlinked, which is characterized by the Berry phase distribution and the surface flat bands \cite{SM}. The properties of the surface states for different configurations of nodal loops and trivially-linked nodal loops were also discussed in the Supplemental Material \cite{SM}.

\emph{Experimental realization  with cold atoms  in optical lattices}.---The high controllability of ultracold atoms in optical lattices makes them a suitable platform for the investigation of exotic physics \cite{Lewenstein}. Many novel techniques have been developed for use with these systems, such as laser-assisted tunneling \cite{Miyake}, optical lattice shaking \cite{Struck}, Raman transition tunneling \cite{Wu}, atomic interferometry \cite{Atala,Duca}, and Bragg scattering \cite{Stamper-Kurn}. These techniques can be used to emulate  physical phenomena that are difficult to realize in solid materials.  Here, we present a scheme  to realize and detect the topological semimetal with DHNL using cold atoms in an   optical lattice. A tight-binding Hamiltonian can be defined as
\begin{eqnarray}
H&=&-t_1\sum_i [c_i^\dag\sigma_z c_{i+\hat{x}}+c_i^\dag\sigma_z c_{i+\hat{y}}]+H.c.\nonumber\\
&&+t_2\sum_i [e^{-\frac{\pi}{2}i}c_i^\dag\sigma_xc_{i+\hat{y}+\hat{z}}+e^{-\frac{\pi}{2}i}c_i^\dag\sigma_xc_{i+\hat{y}-\hat{z}}]+H.c.\nonumber \\
&&+t_2\sum_i [c_i^\dag\sigma_x c_{i+\hat{x}+\hat{z}}-c_i^\dag\sigma_x c_{i+\hat{x}-\hat{z}}]+H.c.\nonumber\\
&&+ {\chi}\sum_i  c_i^\dag \sigma_z c_i\label{latt}
\end{eqnarray}
where the lattice constant is set to unity, and $\hat{x}, \hat{y}, \hat{z}$ are the primitive lattice vectors. The operators $c_{i,\sigma}$ are defined in the Wannier representation $w_i(\bm{r})$ at the lattice site $i$, with pseudospin $\sigma=\uparrow,\downarrow$ representing two intrinsic atomic states. The Bloch Hamiltonian (\ref{H}) can be recovered by performing a Fourier transformation on Eq. (\ref{latt}), and setting the hopping coefficients to $t_1=1$, and $t_2=1/4$.

To achieve the Hamiltonian (\ref{latt}), we chose two hyperfine spin states $|1,-1\rangle$ and $|1,0\rangle$ of $^{87}$Rb as the pseudospins in our model, and constructed a spin-dependent cubic optical lattice using several lasers to trap two pseudospin atoms.  The lattice potential along the $z$ direction was sufficiently deep, such that the hopping along the $z$ direction was negligibly weak \cite{note}. The optical lattice shaking technique was applied to the spin-down optical lattice, which resulted in the renormalization of the hopping coefficient of the spin-down atoms with a negative sign \cite{Struck}. Hence, the hopping coefficients of the spin-up and spin-down atoms in the $x$-$y$ plane had opposite signs. Equal hopping strengths for both spins can be achieved by fine-tuning the depths of the two optical lattices. The Zeeman term  in Eq. (\ref{latt}) can be constructed and  tuned by applying an external magnetic field along the $z$ direction.
The diagonal spin-flip hopping in the $y$-$z$ and $x$-$z$ planes, as shown in the second and third lines of Eq. (\ref{latt}), can be achieved using two groups of  Raman fields \cite{SM}. By properly designing the Raman lattices, onsite and nearest-neighbor spin-flip hopping are forbidden. The hopping-accompanying  phase in the $y$-$z$ plane can be achieved by tilting a pair of Raman lasers at an appropriate angle from the $y$ axis in the $y$-$z$ plane \cite{SM}.

The configuration of the DHNL and the attached Berry flux can be measured using an interferometric technique with high momentum resolution \cite{Atala,Duca}. The energy band of the surface states can be probed by Bragg spectroscopy \cite{Stamper-Kurn}.

\emph{Note added.}---After submitting our manuscript for review, we became aware of the works by Yan \emph{et al.} \cite{Yan}.

\begin{acknowledgments}
We acknowledge helpful discussions with Y. X. Zhao, D. W. Zhang, W. Y. Deng, F. Mei and Z. Y. Xue. This work was supported by the National Natural Science Foundation of China under Grants No. 11504171 (W.C.) and No. 11274061 (J.M.H.). W.C. was also supported by the Natural Science Foundation of Jiangsu Province in China under Grant No. BK20150734. H.-Z.L. is supported by Guangdong Innovative and Entrepreneurial Research Team Program (Grant No. 2016ZT06D348), the National Key R \& D Program (Grant No. 2016YFA0301700), and National Natural Science Foundation of China (Grant No. 11574127).
\end{acknowledgments}

\newpage
\onecolumngrid
\renewcommand{\theequation}{S.\arabic{equation}}
\setcounter{equation}{0}
\renewcommand{\thefigure}{S.\arabic{figure}}
\setcounter{figure}{0}

\section{Supplemental Material for ``Topological semimetals with double-helix nodal link''}

\subsection{Derivation of the Hamiltonian from the method of Hopf map}
In general, a two-band Hamiltonian (trivial energy shift term neglected) in three dimensions can be written as
\begin{eqnarray}
\mathcal{H}(\bi{k})=d_x(\bi{k})\sigma_x+d_y(\bi{k})\sigma_y+d_z(\bi{k})\sigma_z
\end{eqnarray}
$(\hat{d}_x(\bi{k}), \hat{d}_y(\bi{k}), \hat{d}_z(\bi{k}))=( {d}_x(\bi{k}),  {d}_y(\bi{k}),  {d}_z(\bi{k}))/|\bi{d}(\bi{k})|$ is actually a mapping from $T^3$ to $S^2$, which is characterized by the Hopf number. In order to construct such a map, we define a $CP^1$ field as
$z(\bi{k})=(z_1(\bi{k}), z_2(\bi{k}))^T$ with  $z_1(\bi{k})=N_1(\bi{k})+iN_2(\bi{k}), z_2(\bi{k})=N_3(\bi{k})+iN_4(\bi{k})$, where
 \begin{eqnarray*}
&&N_1(\bi{k})=\cos k_z, N_2(\bi{k})=\sin k_z\\
&&N_3(\bi{k})  = -\frac{1}{2}(\sin k_x \cos k_z \sin k_z- \sin k_y \cos^2 k_z)
+\frac{1}{2}\sin k_z\sqrt{(8\cos k_x+8\cos k_y+4\chi+4) - ( \sin k_x\sin  k_z-\sin  k_y \cos  k_z )^2}\\
&&N_4(\bi{k}) =-\frac{1}{2}(\sin k_x  \sin^2 k_z- \sin k_y \cos k_z\sin k_z)
-\frac{1}{2}\cos k_z\sqrt{(8\cos k_x+8\cos k_y+4\chi+4) - ( \sin  k_x\sin  k_z-\sin  k_y \cos k_z )^2}
\end{eqnarray*}
The Hopf maps can be classified by the  Hopf number defined as \cite{Moore,Deng1,Deng2,Kennedy,Liu}
\begin{eqnarray}
\Gamma=\frac{1}{2\pi^2}\int_{BZ} d\bi{k}\epsilon_{abcd}\hat{N}_a\partial_{k_x}\hat{N}_b\partial_{k_y}\hat{N}_c\partial_{k_z}\hat{N}_d
\end{eqnarray}
where $\hat{N}_i=N_i/\sqrt{\sum_{j=1}^4N_j^2}$.
In order to derive our Hamiltonian, we define $d_i(\bi{k})=z^\dag(\bi{k}) \sigma_i z(\bi{k})$, then we have
\begin{eqnarray}
d_x(\mathbf{k})&=&\sin k_y\cos k_z-\sin k_x\sin k_z\\
d_y(\mathbf{k})&=&-\sin(2k_z)\sqrt{(8\cos k_x+8\cos k_y+4\chi+4) - ( \sin  k_x\sin  k_z-\sin  k_y \cos k_z )^2}\\
d_z(\mathbf{k})&=&2\cos k_x+2\cos k_y+\chi
\end{eqnarray}
For a system with PT symmetry, the Hamiltonian can be written as
\begin{eqnarray}
\mathcal{H}_{PT}(\bi{k})=d_x(\bi{k})\sigma_x+d_z(\bi{k})\sigma_z
\end{eqnarray}
where the $\sigma_y$ term is forbidden by PT symmetry. For the Fermi surface, it is required to satisfy
\begin{eqnarray}
d_x(\bi{k})=d_z(\bi{k})=0
\label{nl}
\end{eqnarray}
It is easy to verify that the solution of Eq.(\ref{nl}) should be   loops in the 3D Brillouin zone.
From the Hopf map, one can find that these loops may map into two points $(0,\pm 1,0)$ on $S^2$. Therefore, there exists two possible loops mapping into two points, respectively.  The link relation of two loops can be  characterized by the Hopf number $\Gamma$.

\subsection{Berry phase integral in the $x$ direction and surface states}

\begin{figure}
\centering
\includegraphics[width=0.7\textwidth]{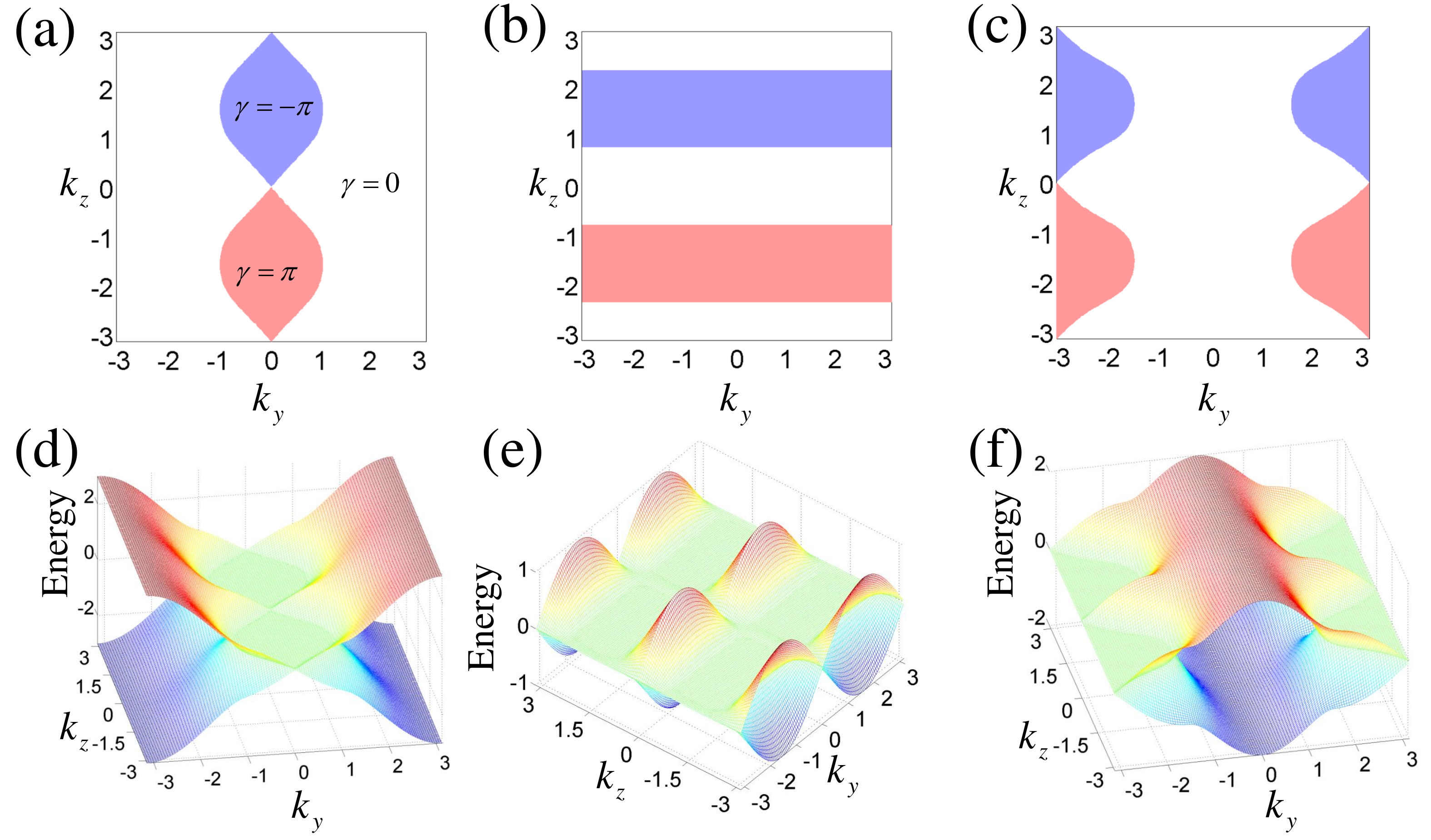}
\caption{(Color online). (a)-(c) Berry phase distributions in the projected surface Brillouin zone as a function of $k_y$ and $k_z$, corresponding to $\chi=-3, 0$, and $2$, respectively. (d)-(f) Corresponding surface states on the open boundaries in the $x$ direction.
} \label{figs1}
\end{figure}

We chose the integral path $\ell$ in Eq. (7) of the main text to be parallel to the $x$ axis. Hence, the Berry phase can be calculated by
\begin{eqnarray}
\gamma&=&i\int_{-\pi}^\pi dk_x\langle u_-(\bm{k})|\partial_{k_x}|u_-(\bm{k})\rangle\nonumber\\
&=&\frac{i}{2}\int_{-\pi}^\pi dk_x\big[\partial_{k_x}\ln \text{sgn}(d_x)\big](1-\cos\theta),
\end{eqnarray}
which can be further reduced as below:
\begin{eqnarray}
\gamma=\frac{i}{2}\sum_{j=1,2}\int_{k'_j-\delta}^{k'_j+\delta}dk_x\big[\partial_{k_x}\ln \text{sgn}(d_x)\big]\nonumber\\
\times\big[1+\text{sgn}(2\cos k_y+2\cos k'_j+\chi)\big],
\end{eqnarray}
where $k_x=k'_1,k'_2$ are the two real roots of the equation $\sin k_x=\sin k_y\cot k_z$. The distributions of the Berry phase in the projected surface Brillouin zone for different values of $\chi$ are shown in Fig. \ref{figs1}(a)-(c). The corresponding surface states are presented in Fig. \ref{figs1}(d)-(f), where the flat bands indicate the region with a nontrivial Berry phase.

\subsection{Unlinking of the double-helix nodal link}
\begin{figure}
\centering
\includegraphics[width=0.7\textwidth]{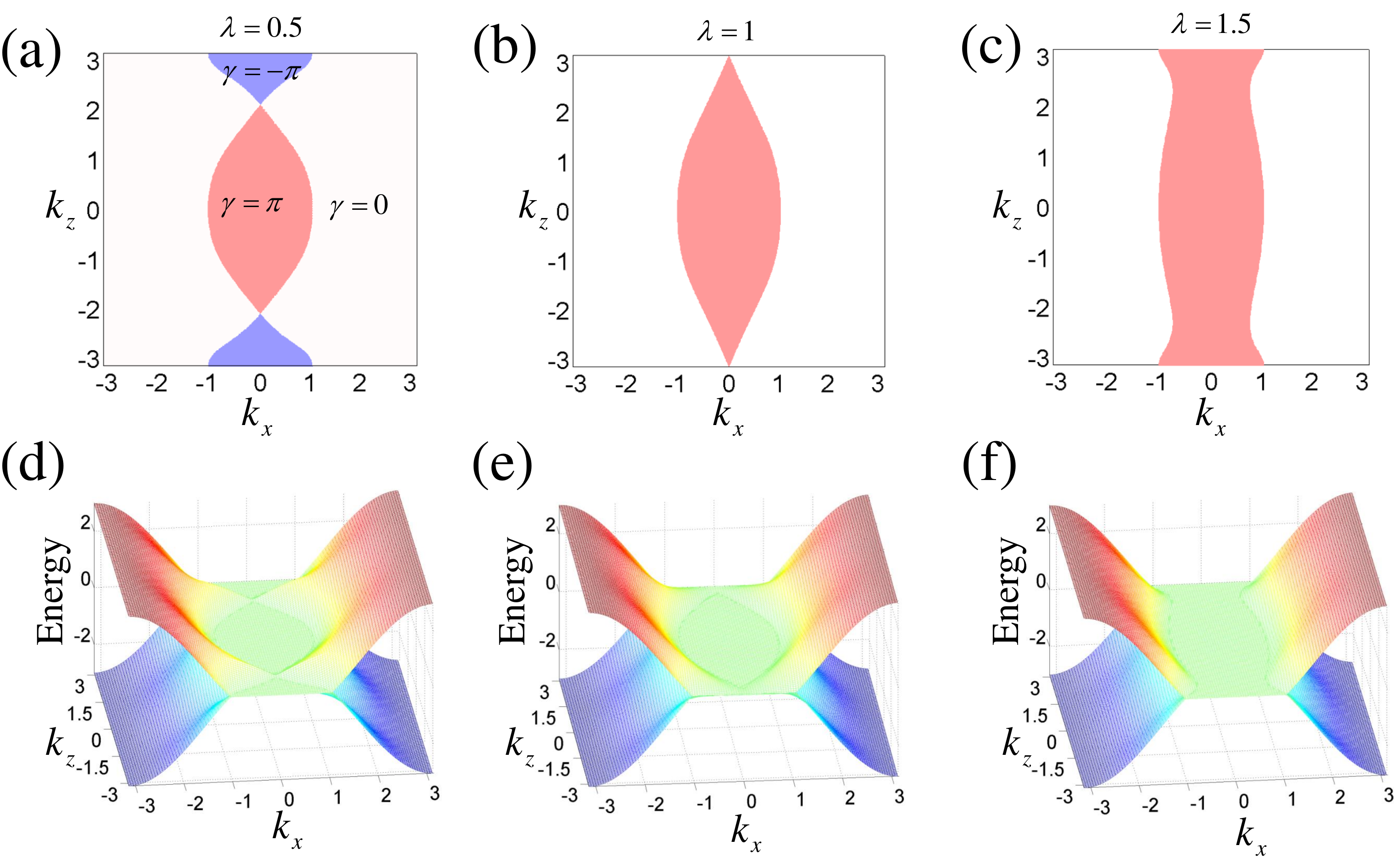}
\caption{(Color online). (a)-(c) Berry phase distributions in the projected surface Brillouin zone as a function of $k_x$ and $k_z$, corresponding to $\lambda=0.5, 1$, and $1.5$, respectively. (d)-(f) Corresponding surface states on the open boundaries in the $y$ direction.
} \label{figs11}
\end{figure}

In order to investigate the linking-unlinking transition of the nodal lines, an additional term $\lambda \sin k_y$ is introduced into $d_x$ in Eq. (1) of the main text. The physical meaning of this term is the nearest-neighbor spin-flip hopping along the $y$ direction. The Berry phase distributions and the corresponding surface states as a function of $k_x$ and $k_z$ are shown in Fig. \ref{figs11}. As $\lambda<1$, two nodal lines are linked, so that the projection of them possesses two intersection points, as shown in Fig. \ref{figs11}(a) and (d). When $\lambda$ reaches 1, the projection of the nodal lines touch at a single point, as shown in Fig. \ref{figs11}(b) and (e). As $\lambda>1$, the two nodal lines are unlined, and they are topologically equivalent to two isolated loops. As a result, the projection of them has no intersection point, and the surface flat band is topologically equivalent to a Corbino disk because of the periodicity of the Brillouin zone, as shown in Fig. \ref{figs11}(c) and (f).

\subsection{Discussion on the surface states for different configurations of nodal loops and trivially-linked nodal loops}

\begin{figure}
\centering
\includegraphics[width=0.7\textwidth]{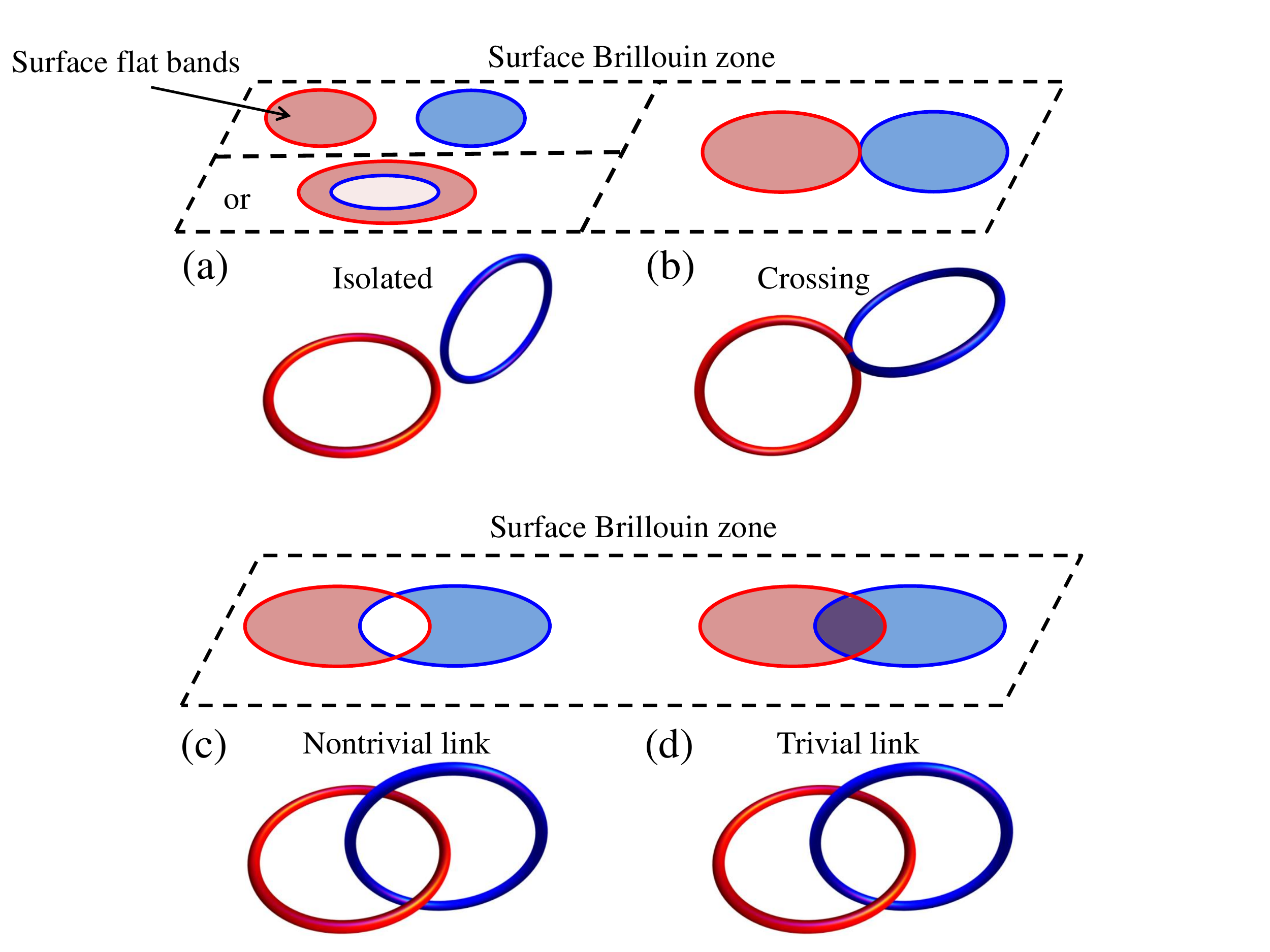}
\caption{(Color online).  Schematic figure of the nodal loops and the surface flat bands in the projected surface Brillouin zone. The nodal loops can be (a) isolated, (b) crossing, (c) nontrivially linked and (d) trivially linked.
} \label{figs4}
\end{figure}

The configurations of the nodal loops can be well reflected by the surface states as shown in the Fig. \ref{figs4}. For a single nodal loop carrying nontrivial Berry phase, it results in surface flat bands nestled inside its projection in the surface Brillouin zone. When there are two nodal loops, the projection of them depends on the surface orientation. The difference between isolated, crossing and linked nodal loops is that, for isolated and crossing ones, there exists a surface orientation, in which the projection of the nodal loops are isolated or touching at a single point; while for the linked ones, the number of intersection points is always two. As a result, the least number of the intersection points of surface flat bands is 0, 1 and 2 for the isolated, crossing and linked cases as shown in the Fig. \ref{figs4} (a), (b) and (c).

In the nontrivially linked case of our work, the surface flat bands disappear in the intersection area of the projection of the nodal loops as shown in Fig. \ref{figs4} (c). For the trivially linked case, two nodal loops can be tuned independently, which come from two decoupled sets of bands, so that the surface flat bands become doubly degenerate in the intersection area as shown in Fig. \ref{figs4} (d).

\subsection{Experimental realization using cold atoms in optical lattices}

 The tight-binding Hamiltonian  (9) in the main text can be written  as $H=H_{xy}+H_{xz}+H_{yz}+H_{Zeeman}$, with
\begin{equation}
\begin{split}
&H_{xy}=-t_1\sum_i[ c_i^\dag\sigma_z c_{i+\hat{x}}+c_i^\dag\sigma_z c_{i+\hat{y}}]+H.c.\\
&H_{yz}=t_2\sum_i [e^{-\frac{\pi}{2}i}c_i^\dag\sigma_xc_{i+\hat{y}+\hat{z}}+e^{-\frac{\pi}{2}i}c_i^\dag\sigma_xc_{i+\hat{y}-\hat{z}}]+H.c. \\
&H_{xz}=t_2\sum_i[ c_i^\dag\sigma_x c_{i+\hat{x}+\hat{z}}-c_i^\dag\sigma_x c_{i+\hat{x}-\hat{z}}]+H.c..\\
&H_{Zeeman}= {\chi} \sum_{i} c_i^\dag \sigma_z c_i
\end{split}
\end{equation}
 \begin{figure}
\centering
\includegraphics[width=0.5\textwidth]{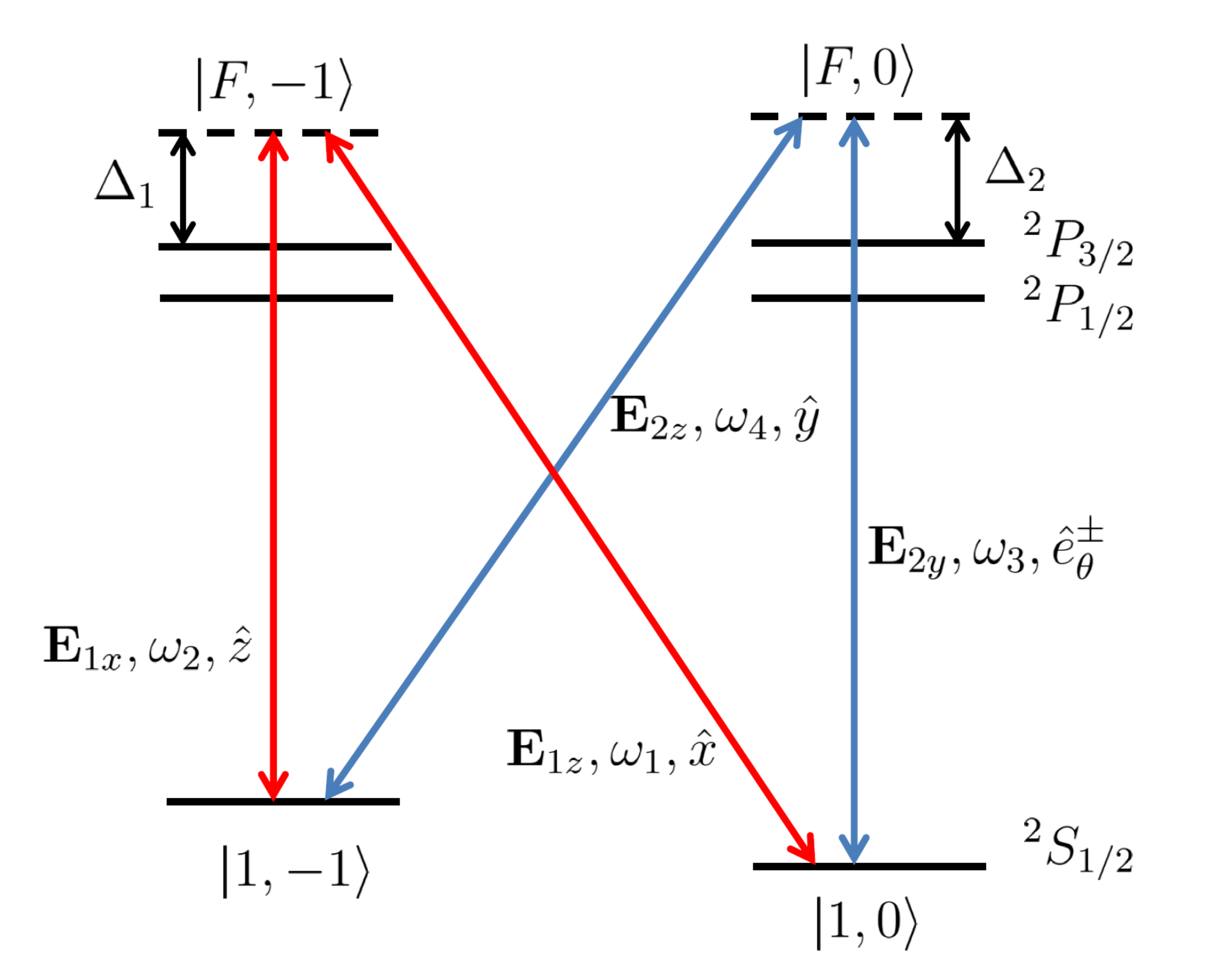}
 \caption{(Color online). Atomic levels of $^{87}$Rb and the Raman lasers used in the present study.} \label{figs2}
\end{figure}

\em Basic experimental setup\em : In our model, we chose two hyperfine spin states $|1,-1\rangle$ and $|1,0\rangle$  of $^{87}$Rb as the pseudospin states $|\uparrow\rangle$ and $|\downarrow\rangle$ \cite{Wu2}. Cold atoms $^{87}$Rb were trapped in a spin-dependent cubic optical lattice   with a lattice potential of $V_\sigma(\bm{r})=-V_{1,\sigma}\big [\cos^2(\frac{\pi x}{a_x})+\cos^2(\frac{\pi y}{a_y})]-V_{2,\sigma}\cos^2(\frac{\pi z}{a_z})$, where $a_{x,y,z}$ are the lattice constants along the $x,y,z$ directions, which can be constructed using several pairs of lasers.

\em Construction of $H_{Zeeman}$\em : The Zeeman  term $H_{Zeeman}$ can be  constructed and tuned by applying an external magnetic field along the $z$ direction, which results in an energy difference of $2\chi$ between the spin-up and spin-down states.

\em Construction of $H_{xy}$\em : The lattice potential along the $z$ direction is chosen such that it is deep enough that hopping along the $z$ direction is negligibly weak, and thus the hopping term along the $z$ direction  can be neglected and only the nearest-neighbor hopping terms in the $x$-$y$ plane are allowed. The corresponding hopping strength in the $x$-$y$ plane  can be written as $t_{1,\sigma}=\int dr^3w_i^*(\bm{r})\big[\bm{p}^2/(2m)+V_\sigma(\bm{r})\big]w_{i+\hat{x}(\hat{y})}(\bm{r})$, where $w_i(\bm{r})$ is a Wannier function centered at site $i$. The lattice shaking technique can be applied to the spin-down optical lattice, which results in the renormalization of the hopping coefficient  of spin-down atoms with a negative sign, such as $t_{1,\downarrow}'=-\kappa t_{1,\downarrow}$, where $\kappa$ is a positive renormalization coefficient\cite{Struck2}. The renormalization coefficient can be fined-tuned by properly choosing the depths of the spin dependent optical lattices, such that $t_{1,\downarrow}'=-t_{1,\uparrow}=-t_1$ in $H_{xy}$.

\begin{figure}
\centering
\includegraphics[width=0.3\textwidth]{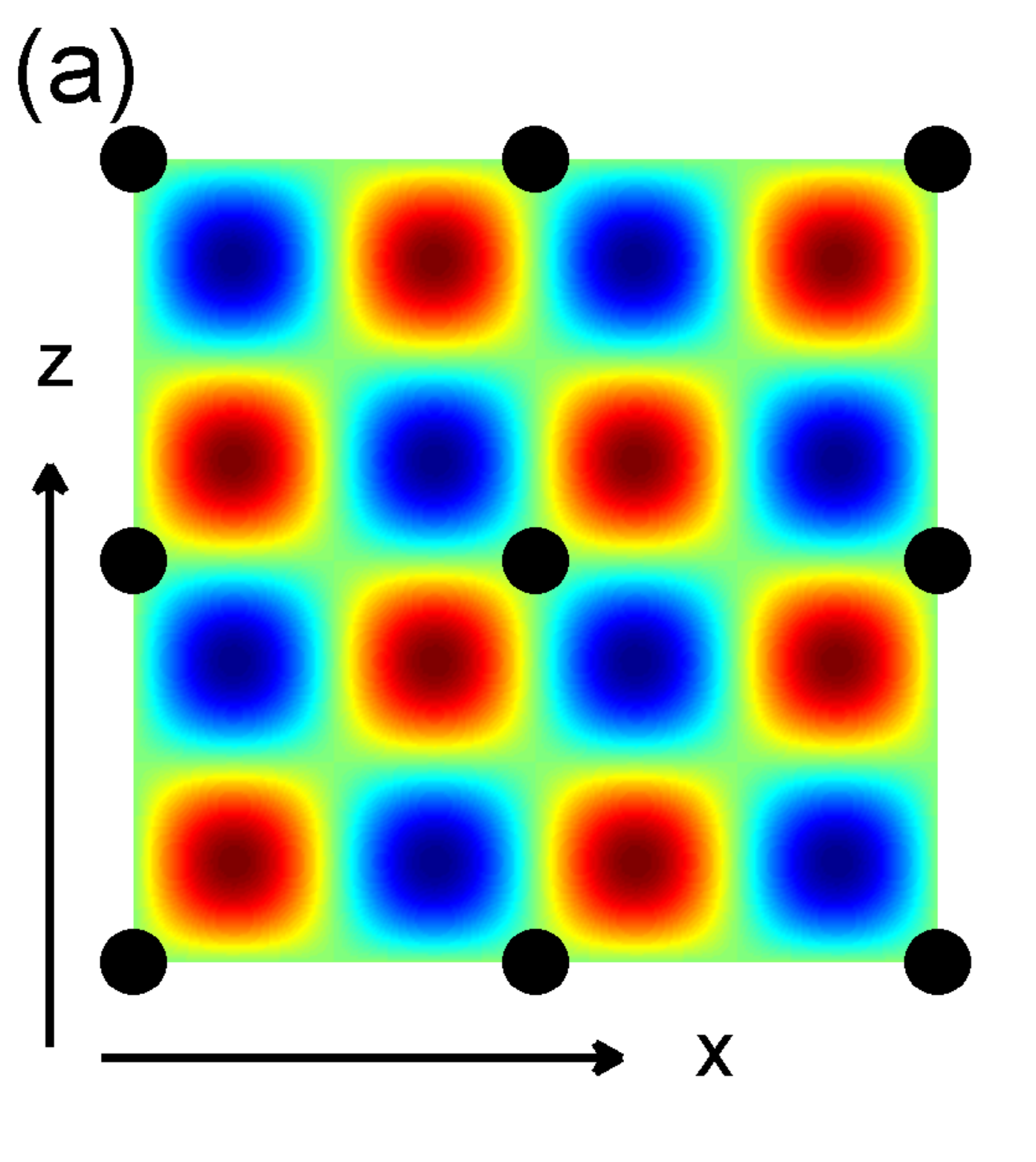}
\includegraphics[width=0.3\textwidth]{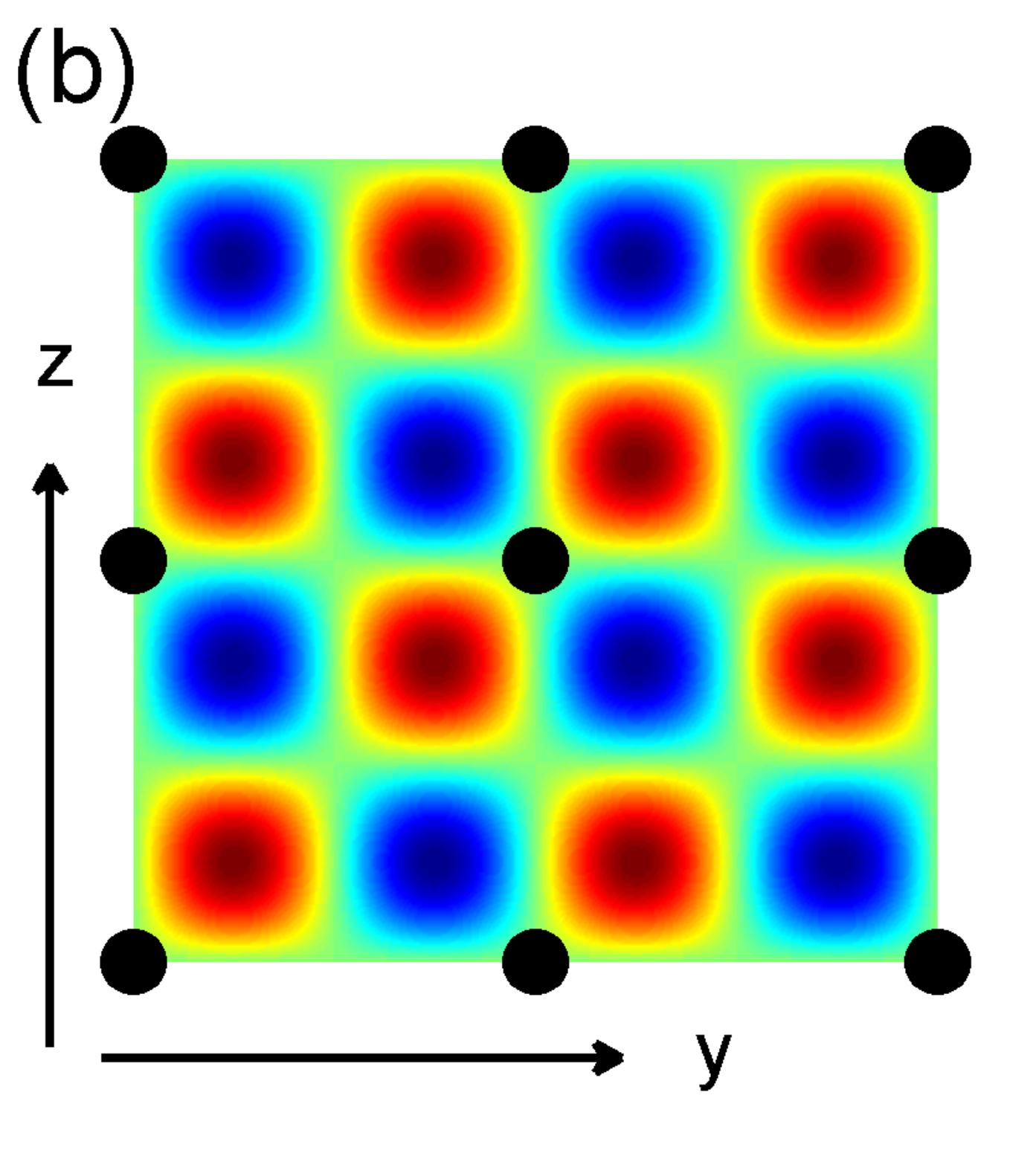}
\caption{(Color online). (a) Profile of the Raman potential $M_1(\bm{r})$ in the $x$-$z$ plane. (b) Intensity of the Raman potential $M_2(\bm{r})$ in the $y$-$z$ plane. The filled circles represent the lattice sites of the optical lattice. } \label{figs3}
\end{figure}

\em Construction of $H_{xz}$\em : $H_{xz}$ describes the diagonal spin-flip hopping in the $x$-$z$ plane, which can be constructed using Raman transitions, as shown in Fig.\ref{figs2}. We chose the excited state $|F,-1\rangle$ as the intermediate state, which provides a Raman transition channel between the ground  states $|1,-1\rangle$ and $|1, 0\rangle$ using two Raman laser fields.  One of the Raman fields is  a standing wave ${\bm E}_{1z}=\hat{x}\mathcal{E}_1\sin(k_1z)$ formed by a pair of oppositely propagating lasers in the $z$ direction, with a frequency of $\omega_1=ck_1$ (where $c$ is the speed of light in vacuum)  and exhibits linear  polarization along the $x$ axis. This field induces the transition between the $|1,0\rangle$ and $|F,-1\rangle$ states. The other Raman laser field is a standing wave ${\bm E}_{1x}=\hat{z}\mathcal{E}_1\sin(k_2x)$  formed by a pair of oppositely propagating lasers in the $x$ direction, with a frequency of $\omega_2=ck_2$ and linear polarization along the $z$ axis. This field results in the transition between states $|1,-1\rangle$ and $|F,-1\rangle$. To avoid pumping atoms to the intermediate state, the frequencies $\omega_1$ and $\omega_2$ were chosen to have a large detuning $\Delta_1$ from the energy differences between the ground states and the intermediate state. The frequency difference $\delta\omega=\omega_1-\omega_2$ was set to match the Zeeman splitting between the $|1,-1\rangle$ and $|1,0\rangle$ states.  We set $k_1=2\pi/a_z$ and $k_2=2\pi/a_x$ by fine-tuning the frequencies of the Raman lasers to ensure that the Raman lattice matches the optical lattice, as shown in Fig.\ref{figs3}(a). The Raman potential, which induces spin-flip transitions, takes the form of $M_1(\bm{r})=\mathcal{M}_1\sin(2\pi x/a_x)\sin(2\pi z/a_z)$, where $\mathcal{M}_1\propto \mathcal{E}_1^2$\cite{Wu2}. The profile of the Raman potential  in the $x$-$z$ plane is shown in Fig.\ref{figs3}(a). The Raman potential has opposite signs along the two diagonal directions, such that the amplitude of the on-site spin-flip transition vanishes and the nearest-neighbor spin-flip hopping terms in the $x$-$z$ plane are   forbidden. However, the Raman potential induces spin-flip hopping along the diagonal directions in the $x$-$z$ plane. The hopping coefficients of the two diagonal hopping terms have opposite signs, such that $t_{xz}=\int dr^3w_i^*(\bm{r})M_1(\bm{r})w_{i+\hat{x}+\hat{z}}(\bm{r})=-\int dr^3w_i^*(\bm{r})M_1(\bm{r})w_{i+\hat{x}-\hat{z}}(\bm{r})$. $t_{xz}=t_2$ in $H_{xz}$ can be achieved by tuning the amplitude $\mathcal{E}_1$ of the Raman lasers.

\em Construction of $H_{yz}$\em : $H_{yz}$ describes the diagonal spin-flip hopping with an accompanying phase  in the $y$-$z$ plane, which can be constructed by an additional Raman transition channel via the intermediate state $|F, 0\rangle$, as shown in Fig.\ref{figs2}.  One of the Raman laser fields is a standing wave $\bm{E}_{2z}=\hat{y}\mathcal{E}_2\sin(k_4z)$ formed by a pair of lasers oppositely propagating in the $z$ direction, with a frequency of $\omega_4=ck_4$ and liner polarization along  the $y$ axis. This field induces the transition between states $|1,-1\rangle$ and $|F,0\rangle$. The other Raman laser field is  formed by two non-collinear lasers propagating in the $y$-$z$ plane, with frequency $\omega_3=ck_3$. The two lasers are described by $E_{2y}^{\pm}=\frac{1}{2}(\hat{z}\cos\theta\pm\hat{y}\sin\theta )\mathcal{E}_2e^{\pm i[(k_3y\cos\theta\mp k_3z\sin\theta)-\pi/2]}$, where $\theta$ is the included angle between the lasers and the $y$ axis. The two lasers form the field $E_{2y}=E_{2y}^++E_{2y}^-=E_{2y}^{(z)}+E_{2y}^{(y)}$, where $E_{2y}^{(z)}=\hat{z}\cos\theta\mathcal{E}_2\sin(k_3y\cos\theta)e^{-ik_3z\sin\theta}$ and $E_{2y}^{(y)}=\hat{y}\sin\theta \mathcal{E}_2\cos(k_3y\cos\theta)e^{-i(k_3z\sin\theta+\pi/2)}$. The laser field $E_{2y}^{(z)}$ results in the transition between states $|1,0\rangle$ and $|F,0\rangle$. To avoid pumping atoms to the intermediate state, the frequencies $\omega_3$ and $\omega_4$ were chosen to exhibit a large detuning $\Delta_2$ from the energy differences between the ground states and the intermediate state.  The Raman potential is $M_2(\bm{r})=\mathcal{M}_2\sin(k_3y\cos\theta)\sin(k_4z)e^{-ik_3z\sin\theta}$, where $\mathcal{M}_2\propto \mathcal{E}_2^2 \cos\theta$\cite{Wu2}. The hopping-accompanying phases can be achieved by adjusting the angle $\theta$ to satisfy the equality $k_3\sin\theta =\pi/2a_z$. To ensure that the optical lattice matches with the Raman lattice, we set $k_3\cos\theta =2\pi/a_y$ by tuning the lattice constant $a_y$ in the $y$ direction. The difference between $\omega_1$ and $\omega_4$ is significantly less than $\omega_1$ and $\omega_4$. Therefore, the difference between $k_1$ and $k_4$ can be neglected, which gives $k_4=2\pi/a_z$. The Raman potential can be rewritten as $M_2(\bm{r})=\mathcal{M}_2\sin(2\pi y/a_y)\sin(2\pi z/a_z)e^{-i\pi z/2a_z}$. The intensity of $M_2(\bm{r})$ in the $y$-$z$ plane is shown in Fig.\ref{figs3}(b), which indicates that the amplitude of the on-site spin-flip transition is vanishing and the nearest-neighbor spin-flip hopping terms in the $y$-$z$ plane are negligible.  The diagonal hopping coefficients  in the $y$-$z$ plane are $t_{yz}e^{-i\pi/2}=\int dr^3w_i^*(\bm{r})M_2(\bm{r})e^{-i\pi z/2a_z}w_{i+\hat{y}+\hat{z}}(\bm{r})= \int dr^3w_i^*(\bm{r})M_2(\bm{r})e^{-i\pi z/2a_z}w_{i+\hat{y}-\hat{z}}(\bm{r})$. $t_{yz}=t_2$ in $H_{yz}$ can be achieved by tuning the amplitude $\mathcal{E}_2$ of the Raman lasers.


\begin{thebibliography}{99}
\bibitem{note} For simplicity, we have omitted the hopping term along the $z$ direction in Eq. (\ref{latt}), because the hopping coefficient is much smaller than $t_1$. It is easy to verify that weak hopping along the $z$ direction does not change the configurations or the Berry flux of the nodal lines.
\end{thebibliography}
\end{document}